\documentclass[proof]{WileyASNA-v1}

\articletype{Original article}%

\received{xx xxxxx xxxx}
\revised{xx xxxxx xxxx}
\accepted{xx xxxxx xxxx}

\begin{document}

\title{Review of light curves of novae in the modified scales. 
	III. V1047 Cen at 2019.}

\author[1]{Rosenbush A.*}




\address[1]{\orgname{Main Astronomical Observatory of the National Academy of Sciences of Ukraine}, \orgaddress{\state{27 Akademika Zabolotnoho St., 03143 Kyiv}, \country{Ukraine}}}



\corres{*\email{aeros@mao.kiev.ua}}


\abstract{Unique Nova Centauri 2005, V1047 Cen, by a form of the modified light curve was included into the CP Pup group. It means that the re-brightening in 2019 occurred still during the finished stage of the 2005 outburst and can be connected with the current activity of the post-nova. 
	
The light curve of V1047 in 2019 as a separate event is some details similar to the HR Del, Nova Del 1967, and also it is distinguish considerable from dwarf nova outbursts in post-nova systems (GK Per, etc.). 
	
There is proposed possible correct identification of Nova Sgr 1970, V3645 Sgr, which was included into the RR Pic group. 
}


\jnlcitation{\cname{%
\author{Rosenbush A.}} (\cyear{2020}), 
\ctitle{Review of light curves of novae in the modified scales. III. V1047 Cen at 2019.}, \cjournal{Astron. Nachr.}, \cvol{20--;00:1--6}.}


\maketitle


\section{Introduction}\label{sec1}

In the first parts of this study \citep{Rosenbush202Xa, Rosenbush202Xb}, we performed a review of the photometric data of about 500 novae, including classical and recurrent novae, in the active phase. Review was carried out by comparing the light curves displayed in modified scales: the logarithmic scale of a radius of shell ejected during a nova outburst and the magnitude scale. Calibration of each scale was defined by certain rules. The brightness of a nova was normalized to the level of a quiet state, i.e. we were dealing with an outburst amplitude. The abscissa scale was normalized on the moment of maximum brightness. For fast novae, it was a principal light maximum, which was often well defined. For slow novae, as a moment of maximum was considered to be the transition from a rapid increase in brightness to a slower growth before maximum values, which can also be identified with a known pre-maximum halt before the final rise. From this moment, the state of maximum brightness began, the duration of that can be compared with the duration of a nova state with the brightness above a certain level according to \cite{Arp1956}. Since at the maximum brightness there is an ejection of a substance forming an expanding shell, in which dust condensation occurs at a certain radius \citep{Clayton1976} and other processes, assuming that the speed of shell expansion is constant, we switched from the time scale to the shell radius scale in a logarithmic representation. In this case, it could been arose some controlled nuances with the determination of these parameters; therefore, we are forced to refer the first parts of our study for the details of an entire procedure for displaying the light curve in modified scales. An important result of this idea was the stable shape of the light curves at the each stage of the outburst and in the region of transitions from one stage to another, which made it possible to extrapolate the remaining parts of the light curves. An extrapolation of light curve forward for modern novae and confirm the preliminary results as new photometric data become available were for us as the verification of this result. 

The result of our review was a confirmation of our preliminary results of 1999 \citep{Rosenbush1999a, Rosenbush1999b, Rosenbush1999c, Rosenbush2002} about the existence of classical (CN) and recurrent (RN) novae groups with certain forms of light curves. The shape of light curve is determined almost uniquely in the presence of sufficient photometric material. Between groups of novae there is a difference in certain properties: possible dust condensation, geometric shape of discarded shells. The unique V1280 Sco light curve, along with recurrent novae, became the foundation for our research, which simplified the understanding of a light curve comparison process. 

Naturally, there were classical novae with spectral confirmation of an outburst type, but with a unique light curve, presented in 1-2 exemplars \citep{Rosenbush202Xb}. When preparing materials for publication, an object with unique behaviour appeared. His light curve with rare observations before the final brightness decline allowed us to classify him as a nova of the CP Pup group. This object - Nova Cen 2005 - suddenly brightened again at almost 6$^{m}$ in 2019 as a slow nova. By the end of a visibility season of object, it became clear to us that a nova, if viewed separately from the 2005 outburst, may have a light curve typical of the HR Del subgroup of the RR Pic group. We postponed the final decision until the visibility season of 2020 and now we can more confidently draw some conclusions regarding this very unique classical nova.

\section{Nova Cen 2005}\label{sec2} 

Classical Nova Cen 2005, V1047 Cen, was discovered by \cite{Liller2005}. The brightness of this fast nova reached 8.5$^{m}$ with an orange filter. The end of the 2005 visibility season, the weakness of object in the next season and the absence of any peculiar features resulted in the light curve with a small number of observations and a considerable scatter of points. 

Photometric data before the re-brightening of 2019 contributed to the opinion that an outburst of 2005 ended: during the years 2013-2018 V1047 Cen had a mean magnitude of I=17.12${\pm}$0.10$^{m}$ \citep{Mroz2019, Geballe2019}.

\section{Re-brightening of 2019 - rare-observed activity of classical nova during a final brightness decline stage}\label{sec3}

In the 2019 outburst, the light curve of V1047 Cen was radically different from the 2005 one if it is considered as independent event: the amplitude of an outburst was as of a dwarf nova (DN) \citep{Geballe2019}. The DN outbursts of classical novae are not uncommon: the most famous examples are GK Per and V446 Her (two members of the CP Pup group \citep{Rosenbush202Xb}), V1017 Sgr (an outburst in 1917 is the candidate for RN of the T Pyx group \citep{Rosenbush202Xa}). But as noted \cite{Geballe2019}), the 14 years between CN and DN outbursts of V1047 Cen is the shortest of all known and it is a problem for standard theories. According to the discussion of near-infrared spectroscopy they are inclined to believe that it was nevertheless a DN outburst. 

The presentation of 2005 outburst light curve in our modification indicates that the V1047 Cen in 2019 was still at the second half of final brightness decline stage.

We briefly recall that the modified light curve is the behaviour of nova brightness relatively its quiet state, i.e., on the scale of outburst amplitude \citep{Rosenbush1999a, Rosenbush202Xb}, in the dependence on logarithm of a radius of the shell which was ejected during the outburst

log(r)=log(t-t$_{0}$) + C,      ${   }$             (1) 

where C=13.1 is the constant on which the abscissa scale is calibrated, and which is equal to the logarithm of radius r$_{0}$ of the shell at a moment of its ejection from the nova. A time t and a shell radius r in expression (1) are in days and centimetres, respectively. Corresponding parameters for V1047 Cen outbursts in 2005 and 2019 are given in Table 1. Such a representation of the abscissa scale gives simultaneously an idea on the geometric dimensions of the shell in an absence of corresponding spectral observations.

An important parameter for us is the amplitude of outburst. Progenitor V1047 Cen is missing from the SuperCOSMOS catalogue \citep{Geballe2019}, which is 90 per cent completeness to R$\approx$19.5$^{m}$ and B$_{J}$$\approx$20.5$^{m}$ \citep{Hambly2001}. This, in our approach to determining this parameter, provides the basis for accepting the outburst amplitude of V1047 Cen equal to the average amplitude of the group prototypes with which it is possible to find a better match. And from this we can estimate a brightness of V1047 Cen in a quiet state near 20.5$^{m}$ (Table 1). 

The modified V1047 light curve during the 2005 outburst with parameters of Table 1 is nearly identical to the light curve of V476 Cyg, one of the CP Pup group prototypes (Fig.1). I.e. it was an outburst of a classical nova that did not end as of 2020, and to a level of normal brightness, by analogy with V476 Cyg, it is necessary still to weaken by 4-5$^{m}$. At the brightness final decline stage of V1047 Cen, presented before the 2019 re-brightening by the Gaia mission data, the main trend of the nova brightness decline, which is gave way to a outburst of 2019, is clearly visible. The long-term variability with the low amplitude of 0.3-0.4$^{m}$ in the photometric I-band mentioned by \cite{Mroz2019} is also part of this general trend. By this time, the energy distribution in the spectrum of the white dwarf - post-nova - has returned to its original state, and the observed light curve is the re-radiation by the expanding shell of the radiation of central hot source. 

\begin{center}
	\begin{table}[t]%
		\centering
		\caption{Modified light curve parameters of V1047 Cen.\label{tab1}}%
		\tabcolsep=0pt%
		\begin{tabular*}{220pt}{@{\extracolsep\fill}rcc@{\extracolsep\fill}}
			\toprule
			\textbf{Year}& \textbf{Adopted magnitude}  & \textbf{Maximal brightness} \\
			\textbf{}& \textbf{of quiescence, m$_{q}$} & \textbf{moment, t$_{0}$, JD}	\\
			\midrule
			2005    & V=20.5$^{m}$  & 2453614   \\
			2019 & V=20.5$^{m}$    & 2458658    \\
			\bottomrule
		\end{tabular*}
	\end{table}
\end{center}					

\begin{figure}[t]
	\centerline{\includegraphics[width=78mm]{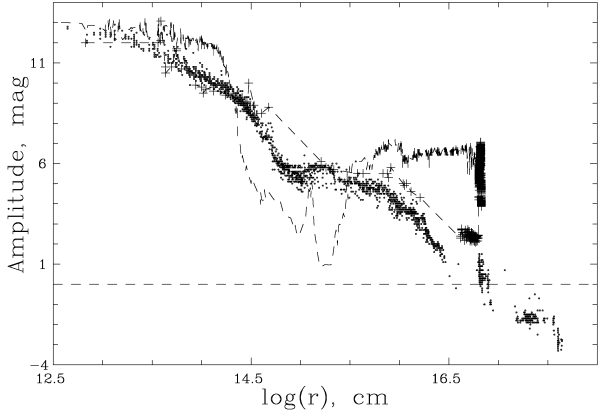}}
	\caption{Complete light curve of V1047 Cen in the 2005 outburst (dashed line with pluses), 476 Cyg (dots) - one of the prototypes of the CP Pup group, and V1280 Sco (dashed line). The amplitude scale is normalized to the average one of the outbursts of 4 prototypes of the CP Pup group \citep{Rosenbush202Xb} (the dashed horizontal line corresponds to the average "null" magnitude of the quiescent of four prototypes of the group). Data from AAVSO, Gaia and \cite{Liller2005} were used.\label{fig1}}
\end{figure}

Belonging to a group implies the presence of common details of light curves as a result of the same physical and others causes of outbursts. Therefore, when comparing with the main prototypes of our classification scheme, we can pay attention to interesting details of the light curve of V1047 Cen. For example, the beginning of 2019 outburst coincides with the completion of a small plateau on the light curves of the prototypes (V476 Cyg, CP Pup) for log(r)$\approx$16.3 (Fig.1). I.e., the central source at this time has a certain activity and this may be a common characteristic of novae in this group. The matter ejected during the principal maximum of outburst, and which is significantly responsible for the visual brightness, is at this time on a distance of about 1000 AU from a central hot source. 

The maximum brightness level of V1047 Cen in the re-brightening was 1$\div$1.5$^{m}$ higher than the typical brightness level of novae in this group at the transition stage of the outburst.

In our review, the unique nova of V1280 Sco \citep{Rosenbush202Xb}, which 13 years after the outburst continues to be at an unusually high level of brightness, played a decisive role. The brightness of V1047 Cen in the re-brightening of 2019 for a short time, on a logarithmic scale, reached that brightness of V1280 Sco. At the same time, the beginning of re-brightening almost coincided with the moment of appearance of quasi-periodic temporal light dips with an amplitude of about 1$^{m}$. 

\section{Discussion}\label{sec4}

Proceeding from one of our starting points \citep{Rosenbush202Xa, Rosenbush202Xb}: the similarity/coincidence of light curves means equal/close physical and geometric parameters of binary systems in which these novae are arose, we can also allow equal outburst amplitudes. I.e., with an average modern visual brightness in the quiet state of V476 Cyg of about 17.4$^{m}$ and the outburst amplitudes of about 15.5$^{m}$ for both novae, we get a possible brightness of a quiet state for V1047 Cen of about 24$^{m}$, which corresponds to the absence of a pre-nova in the SuperCOSMOS catalogue. 2005 outburst of V1047 Cen should end no earlier than 2030, if the brightness of star quiet state is about 20.5$^{m}$; if this brightness is weaker, then the end of outburst is delayed to a later time. Here you can ask the obvious question: what is considered the end of outburst, but without an obvious answer.

Secondary outbursts of novae are not rare. The authors of first detailed study of V1047 Cen \citep{Geballe2019} drew attention to this. The classical nova of GK Per, also a member of the CP Pup group, after the outburst of 1901 showed some more outbursts as a dwarf nova of small amplitude. V1017 Sgr in 1919 showed an outburst in our interpretation \citep{Rosenbush202Xa} typical of the recurrent nova of the T Pyx group, and at least more three DN outbursts were recorded \citep{Salazar2017}. These outbursts occurred a long time after the principal outburst, unlike V1047 Cen with its 14 years between outbursts. An example of a closer relationship between the principal and secondary outbursts is the recurrent nova of T CrB, which in two known cases repeated the secondary re-brightening at the same time after the principal one; the delay time was shorter compared to V1047 Cen. We add that the difference in amplitudes of the primary and secondary outbursts of these two novae is the same 5-5.5$^{m}$. 

The secondary outburst of V1047 Cen, unlike all the other second outbursts of DNs in CNs mentioned above has the highest amplitude of nearly 5$^{m}$ above the current brightness level. But such outburst amplitudes, after our first experiment \citep{Rosenbush1999c}, we are discuss with caution: against the background of small amplitudes of outbursts, it is more difficult to distinguish characteristic details; but outbursts of the so-called “tremendous outburst amplitude dwarf novae” (TOADs) were in our control \citep{Rosenbush202Xb}. The lowest outburst amplitudes among classical novae were among the members of the HR Del of the RR Pic group \citep{Rosenbush202Xb}. 

If the outburst of 2019 to regard as a phenomenon independent of the outburst of 2005, then even at the end of the season of visibility of the sky region with V1047 Cen in 2019, the modified light curve with the parameters of Table 1 was very like on the HR Del light curve (Fig.2), but has not yet reached the final brightness decline stage. This became the basis for us not to include this nova in the second part of our study. With the resumption of observations in 2020, it became apparent that the outburst of 2019 went into the stage of the final brightness decline it is already possible to more definitely compare with the prototypes. 

On Fig.2 our attention attracts immediately the earlier beginning of the final brightness decline stage and the faster brightness decline of V1047 Cen in comparison with HR Del. We tend to consider such differences for low-amplitude outbursts to be sufficient to doubt their identity (a similar difference became the basis for us to form the HR Del subgroup in the RR Pic group). 

By the search for analogues among little-studied novae with small fixed amplitudes of outbursts, for which it was not possible for us to draw certain conclusions, we did not find similar light curves. Here, a large role been played probably the selection of observations due to the low brightness of novae at this stage.

\begin{figure}[t]
	\centerline{\includegraphics[width=78mm]{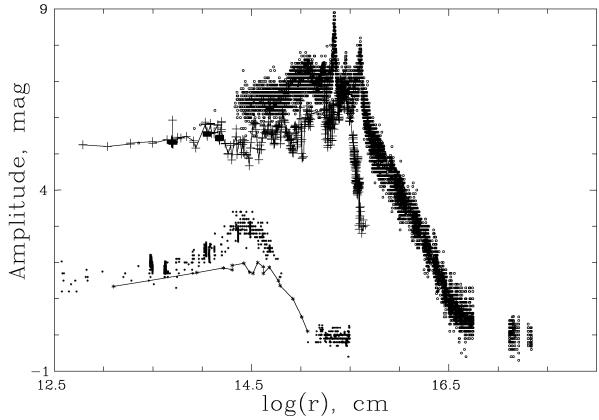}}
	\caption{Light curve of V1047 Cen (line with pluses) during the 2019 “isolated” outburst compared to the HR Del, Nova Del 1967 (dots) light curve \citep{Rosenbush202Xb}. For comparison the modified light curves of dwarf nova outbursts in a post-nova system OGLE-2004-BLG-081 (schematic version, line with dots, \citep{Mroz2015} and GK Per in 2015 (dots, visual data of the AAVSO) are displayed in the left corner.\label{fig2}}
\end{figure}

Here we must understand that the observed light curve during the re-brightening against the background of the final stage of the principal outburst is formed by two sources of visual radiation: a source from the principal outburst, having a brightness of slightly more than 2$^{m}$, and a source associated with the re-brightening and adding a total brightness of up to 7$^{m}$ above the accepted level of quiet state brightness. It is clear that the first source is the shell ejected during the principal outburst. One can only make assumptions about the second source, starting with the repeated ejection of the matter and/or increase in the activity of a post-nova. \cite{Geballe2019} from the spectral line profiles were assumed a possibility, but still were rejected its, that the 2019 profile had generated by two separate outflows. The common emission of these two sources possibly and caused a 1-1.5$^{m}$ excess of the typical V1047 Cen brightness at the transition stage of the outburst, mentioned in Section 3. 

In connection with re-brightening, let us once again draw attention to the recurrent nova of T CrB \citep{Rosenbush202Xa} already mentioned above with the secondary maximum and compare it with the classical nova of the CP Pup group \citep{Rosenbush202Xb}. Fig.3 shows that the secondary maximum of T CrB coincides with the ending of transition stage and the beginning of final stage in the V476 Cyg outburst. [Recall that the abscissa scale in our interpretation is equivalent to the logarithmic time scale according to (1).] On this interval of time/radii, the classical novae develop a nebular spectrum. Interestingly that the depression of RS Oph in the final stage falls within the same interval, i.e. the depression of RS Oph and the secondary maximum of T CrB occur "simultaneously".  

\begin{figure}[t]
	\centerline{\includegraphics[width=78mm]{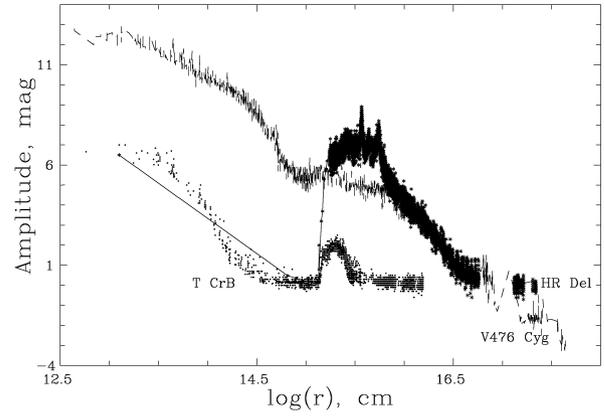}}
	\caption{Light curves of T CrB in 1946 (dots) and V476 Cyg (broken dashed line) compared to the possible hypothetical light curve of HR Del (broken line with dots, a nova name is added next to the final part for the quiet brightness level) during the outburst of 1967 as the secondary to the principal outburst by the T CrB type.\label{fig3}}
\end{figure}

Hypothetically, HR Del, before a recorded outburst of 1967, could have the one like to a RN of the T CrB group \citep{Rosenbush202Xa}. 

We will proceed from the fact that the subgroup of very slow low-amplitude novae as HR Del is very small and contains only 4 members that we confidently identified from the total number of 235 classical novae in our resulting list, or less than 2\%. Let us make the assumption that the observed HR Del outburst in 1967 was a secondary outburst relative to the principal one. Here we consider two options: an outburst of the classical nova and an outburst of recurrent nova.  

Relative to the principal HR Del outburst, a classical nova can be determined after viewing the known photometry until 1967, knowing the brightness of the pre- and post-nova equal to V$\approx$12.1$^{m}$ \citep{Strope2010}. In the interpretation of the classical nova, the principal outburst of HR Del would have occurred in the first half of the 50s of the XX century and the nova would have a maximum brightness of -1$\div$0$^{m}$ and above. These years, more than 10 years, are necessary that the classical nova has time to return to a state close to the initial. Available photometric data \citep{Collazzi2009} exclude such bright state of HR Del. Also the distance to HR Del by the shell expansion parallax d=970$\pm$70 ps \citep{Harman2003} is consistent with the corrected distance by the Gaia mission d=932(+32/-29) ps \citep{Bailer-Jones2018}, i.e. the shell was ejected namely in 1967. Therefore, here could take place the variant with a principal outburst like T CrB and without a shell ejection. The duration of principal outburst was less 50$^{d}$ and it took place during the gap in observations. At this variant we should assume that the secondary maximum was connected with the ejection of a large mass of matter with its concentration in the region of shell "equator" \citep{Harman2003}. 

The parameters of such a outburst: t$_{0}$=JD 2439545, the brightness of a quiet state is 12.1$^{m}$. A possible HR Del light curve with a principal outburst of the type RN of the T CrB group and the secondary as classical is shown in Fig.3 schematically: only maximum for the first and actually observed curve for the secondary. It is noteworthy that in this hypothetical version, the secondary outburst precisely lies at the end of transition stage and completely at the final stage of V476 Cyg up to a small plateau for log(r)$\approx$16.3. 

A comparison of the modified light curves of a dwarf nova outburst in a post-nova system similar to GK Per, etc., and the re-brightening of V1047 Cen shows their significant difference (Fig.2) [this is also evident when comparing traditional light curves]. Earlier we \citep{Rosenbush202Xa, Rosenbush202Xb} came to the conclusion that already a shift of 0.2 to combine two similar in shape curves of light should cause suspicion. Here we have similar in shape light curves, but the abscissa shift, about 1, is very large for the like outburst processes. The outburst amplitude is also less. [We do not give the parameters of modified light curves, since they are not used anywhere in the future. For us, it is important the fundamental difference between the light curves.]

\subsection{V3645 Sgr}

When searching for candidates for the same 2019 outburst as V1047 Cen, we drew attention to the little-studied Nova Sagittarii 1970, V3645 Sgr, discovered from the nebular spectrum on an objective-prism plate taken by R. Bartaja and T.Vashakidse at Abastumani Astrophysical Observatory of Georgia \citep{Arhipova1970}. The light curve was restored from archival images by \cite{Arhipova1971} was added by \cite{Sarajedini1984} (two estimates of the brightness from the spectral images, which form a local short flash in the final part of the light curve, are overestimated, according to the opinion of data authors). The modified light curve (Fig.4) fits well with the prototype of the RR Pic group with parameters t$_{0}$=JD 2440497 and photographic magnitude m$_{q}$=22.3$^{m}$. The maximal magnitude was near 11.8$^{m}$. The brightness of a quiet state that we adopted is at 4.3$^{m}$ faintest from the estimation of \cite{Arhipova1970}. Due to the low accuracy of the coordinates of V3645 Sgr, 1 arc minute, and the comparison chart given of \cite{Arhipova1970}, \cite{Downes2000} recognized the lack the identification of  post-nova, and hence the coordinates of V3645 Sgr. Therefore, the data from Gaia catalogue and its applications \citep{Bailer-Jones2018} is related to a field star. The V3645 Sgr spectrum, which \cite{Surina2013} are described as the spectrum of a cold star, is also belong to a field star. 

\begin{figure}[t]
	\centerline{\includegraphics[width=78mm]{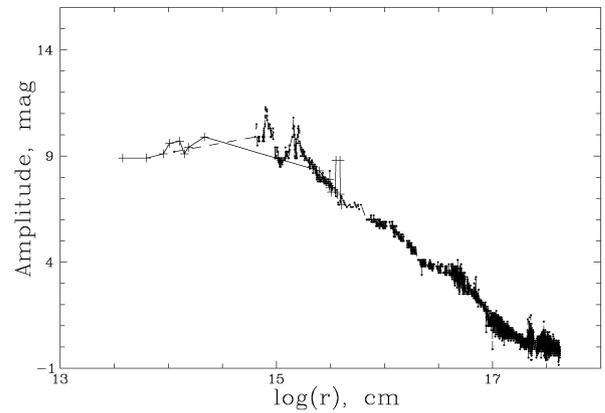}}
	\caption{Modified light curve of V3645 Sgr vs one of RR Pic, the prototype of the group.\label{fig4}}
\end{figure}

If we focus on an object that is present only in the image in the H$\alpha$ filter of \cite{Downes2000}, then the Gaia DR2 4093261823550037376 object with a magnitude of g=20.11$^{m}$ can be the post-nova. Same object is present on the SuperCOSMOS scans \citep{Hambly2001}, which makes it possible to evaluate it as very blue compared to field stars: it is not visible on red but is bright on UKST blue images, and the presence on the red ESO review image can indicate a possible variability.

\section{Conclusion} 

An example of V1047 Cen behaviour in the final stage of the outburst in 2019-2020 draws our attention to the poor knowledge of the processes occurring in the binary system at this stage of the return to a quiescence. At this time, the radiation of expanding envelope remains a good indicator of central source state. The only thing that can slow down the development of investigations in this section of close binary system events is the need for long-term monitoring as these processes developed with the slowed down rate. 

It is necessary also other, more strong criterion to estimate the state of a post-nova as a stability of brightness is not seems as such good criterion. Here can help the coincidence of local activity in objects different types: secondary maximum of RN T CrB and the brightness depression of RN RS Oph on the one hand and the beginning of final brightness decline stage in some CNs, in particular, the CP Pup group, on the other hand. Main difference between these objects is the lack of ejection large mass of matter from the former. 

The inclusion of V3645 Sgr in the RR Pic group confirmed the conclusion of \cite{Downes2000} about the lack of identification for this old nova and made it possible to suggest such identification.

\section*{Acknowledgements}

We thank the AAVSO observers who made the observations on which this project is based, the AAVSO staff who archived them and made publicly available. This research has made use of the NASA's Astrophysics Data System and the SIMBAD database, operated at CDS, Strasbourg, France. This work has made use of data from the European Space Agency (ESA) mission Gaia (https://www.cosmos.esa.int/gaia), processed by the Gaia Data Processing and Analysis Consortium (DPAC, https://www.cosmos.esa.int/web/gaia/dpac/consortium) and the Photometric Science Alerts Team (http://gsaweb.ast.cam.ac.uk/alerts). Funding for the DPAC has been provided by national institutions, in particular the institutions participating in the Gaia Multilateral Agreement. Author is thankful to the Valga vallavalitsus, Estonia, for the support, which allowed us to carry out this interesting investigation.









\nocite{*}
\bibliography{Rosenbush3}%



\end{document}